\documentclass[12pt,psfig,epsfig]{article}
\newcommand {\be}{\begin{equation}}
\newcommand {\ee}{\end{equation}}
\newcommand {\bea}{\begin{eqnarray}}
\newcommand {\eea}{\end{eqnarray}}
\newcommand {\bx}{{\bf x}}
\newcommand {\by}{{\bf y}}
\newcommand {\bX}{{\bf X}}
\newcommand {\bY}{{\bf Y}}

\usepackage{rotating}
\usepackage{epsfig}
\usepackage{psfig}

\begin{document}

\begin{titlepage}

\title{On the performance of different synchronization measures in real
  data: a case study on EEG signals}

\vspace{2cm}
\author{R. Quian Quiroga$^{\dagger *}$, A. Kraskov$^{\dagger}$, 
T. Kreuz$^{\dagger\ddagger}$, and P. Grassberger$^\dagger$\\
$^\dagger${\sl John von Neumann Institute for Computing,}\\
{\sl Forschungszentrum J\"ulich GmbH,}\\
{\sl D - 52425 J\"ulich, Germany}\\
$^\ddagger${\sl Department of Epileptology, University of Bonn,}\\{\sl Sigmund-Freud Str. 25,}\\
{\sl D - 53105 Bonn, Germany} }
\maketitle

\vspace{3cm}
\hspace{0.3cm}   05.45.Xt

\vspace{0.5cm}
$^*$ corresponding author

\newpage
\begin{abstract}
  We study the synchronization between left and right hemisphere rat EEG
  channels by using various synchronization measures, namely
  non-linear interdependences, phase-synchronizations, mutual
  information, cross-correlation and the coherence function.  In
  passing we show a close relation between two recently proposed phase
  synchronization measures and we extend the definition of one of
  them.  In three typical examples we observe that except mutual
  information, all these measures give a useful quantification that is
  hard to be guessed beforehand from the raw data.  Despite their
  differences, results are qualitatively the same.  Therefore, we
  claim that the applied measures are valuable for the study of
  synchronization in real data. Moreover, in the particular case of
  EEG signals their use as complementary variables could be of
  clinical relevance.
\end{abstract}

\end{titlepage}

\newpage
\section{Introduction}

The concept of synchronization goes back to the observation of
interactions between two pendulum clocks by Huygens.  Synchronization
of oscillatory systems has been widely studied but it was not until
recently that synchronization between chaotic motions received
attention. A first push in this direction was the observation of
identical synchronization of chaotic systems
\cite{fujisaka,yamada,pikovskygs,pecora}.  But more interesting has
been the idea of a ``generalized synchronization" relationship as a
mapping between non-identical systems, and the further proposal by
Rulkov et al. \cite{rulkov} of a topological method to quantify it.
The work of Rulkov and coworkers indeed triggered a number of studies
applying the concept of generalized synchronization to real data.  One
of these applications is to the study of electroencephalographic (EEG)
signals, where synchronization phenomena have been increasingly
recognized as a key feature for establishing the communication between
different regions of the brain \cite{singer}, and pathological
synchronization as a main mechanism responsible of an epileptic
seizure \cite{niedermeyer}.  Since many features of EEG signals cannot
be generated by linear models, it is generally argued that non-linear
measures are likely to give more information than the one obtained
with conventional linear approaches.

In a first study dealing with EEG signals, Schiff and coworkers
\cite{schiff} applied a synchronization measure similar to the one
defined in \cite{rulkov} to the study of data from motoneurons within
a spinal cord pool.  More recently, non-linear synchronization
measures were used for the analysis of EEG data from epileptic
patients with the main goal of localizing the epileptogenic zone and
of predicting the seizure onset\cite{vanqyen1,vanqyen2,arnhold}.  These
results, of course, have a clear clinical relevance.  Arnhold and
coworkers \cite{arnhold} proposed a robust measure ($S$), a
variant of which ($H$), already mentioned by these authors, has been
studied in detail in \cite{quian}.  These last two measures of
interdependence, together with a new measure ($N$) to be defined, will
be further studied in this paper.

\begin{figure}
\begin{center}
\epsfig{file=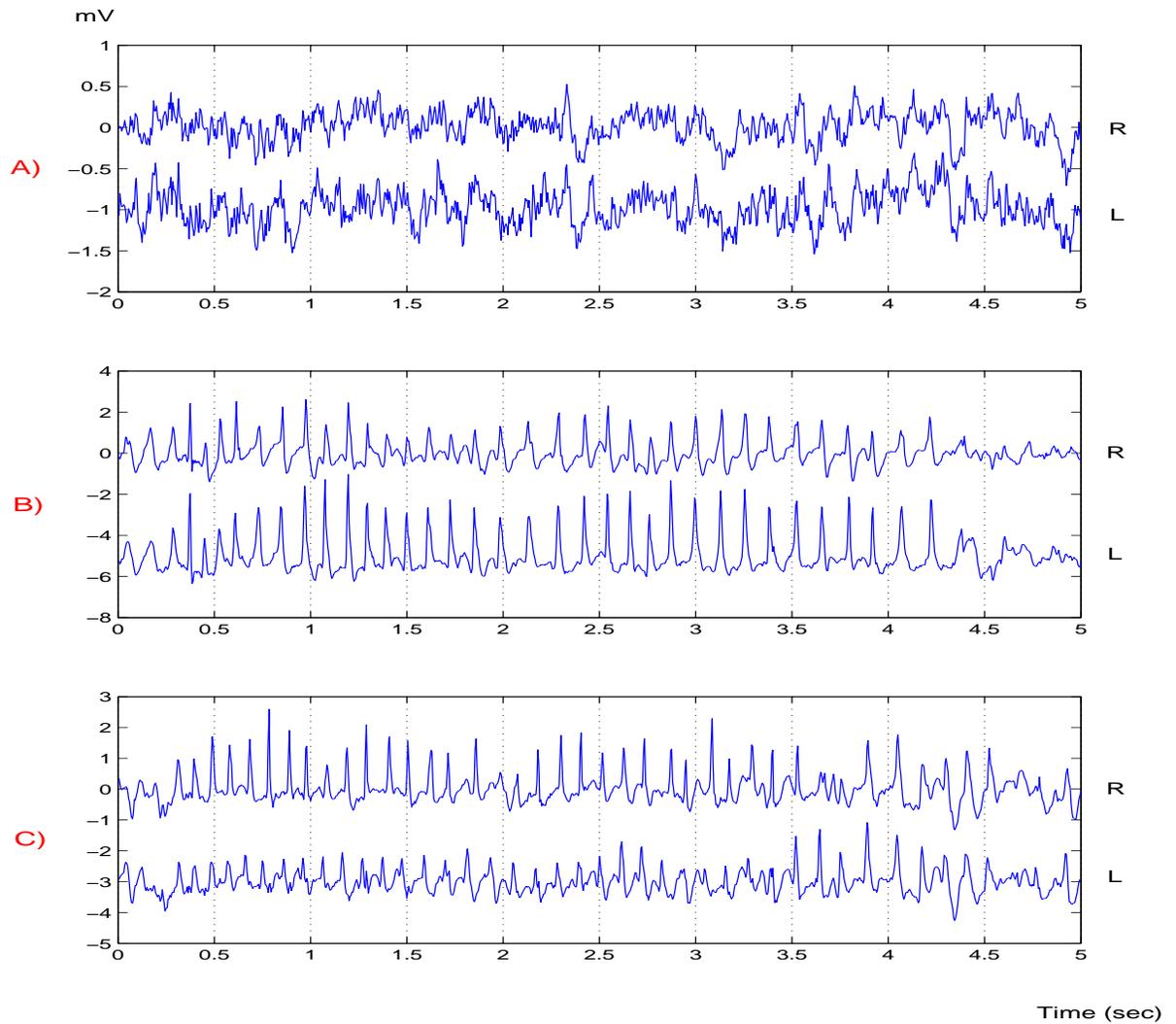, height=14cm,width=16cm,angle=0}
\end{center}
\caption{Three rat EEG signals from right and left cortical
  intracranial electrodes. For a better visualization, left signals
  are plotted with an offset}
\label{fig:examples}
\end{figure}

The previous papers give convincing arguments in favor of using
non-linear interdependences, which in most cases were illustrated with
examples using chaotic toy models.  However, it still remains an open
question whether this also holds true for real data.  In this paper we
therefore address the point of whether non-linear measures give a
relevant contribution to the study of synchronization in
electroencephalographic (EEG) signals \cite{www}.  In particular, we
will show with three typical EEG examples (see Fig.
\ref{fig:examples}) how non-linear interdependence measures can
disclose information difficult to obtain by visual inspection.
Although the data
are EEG recordings from rats, their main features are common
to human EEG.
Moreover, results should not be restricted to EEG data and should also be valuable to
study synchronization of other signals.  For comparison purposes, we
will also study phase synchronization measures as defined from the
Hilbert transform \cite{rosemblum} and from the Wavelet transform
\cite{lachaux}, which had been recently applied to the study of EEG
signals \cite{tass,florian,rodriguez}.  Moreover, we will also compare
these results with the ones obtained with more conventional measures
of synchronization, such as the cross-correlation, the coherence
function and the mutual information.

The paper is organized as follows: In section \ref{sec-measures} we
define the synchronization measures to be used. In particular, in section
\ref{sec-lin} we define the linear cross-correlation and the coherence
function, while in section \ref{sec-h} we describe three recently proposed
measures of non-linear interdependence. The mutual information is
defined in section \ref{sec-mi}, whereas section \ref{sec-ps} is dedicated to
the description of phase synchronization measures with the phases
defined from a Hilbert transform. Very close to these last measures
are the ones described in section \ref{sec-wps} but in this case the
phases are defined from the Wavelet transform. Finally in section
\ref{sec-relation} we show the relation between these two phase
synchronization approaches.  Details of the data sets to be
analyzed are disclosed in section \ref{sec-data}. In section
\ref{sec-results} we describe the results obtained by applying the
different synchronization measures to these data sets. Finally in
section \ref{sec-conclusions} we present the conclusions.

\section{Synchronization measures}
\label{sec-measures}

In the following, unless when further specified, we shall use the
notion of synchronization in a very loose sense. Thus it is more or
less synonymous with interdependence or (strong) correlation.

\subsection{Linear measures of synchronization}
\label{sec-lin}

Let us suppose we have two simultaneously measured discrete univariate
time series $x_n$ and $y_n$, $n=1,\ldots,N$. The most commonly used measure of
their synchronization is the cross-correlation function defined as:

\be
c_{xy}(\tau) = \frac{1}{N-\tau} \sum_{i=1}^{N-\tau} (\frac{x_i -
  \bar{x}}{\sigma_x}) \cdot (\frac{y_{i+\tau} - \bar{y}} {\sigma_y})
\label{eq:cross}
\ee

\noindent
where $\bar{x}$ and $\sigma_{x}$ denote mean and 
variance, and $\tau$ is a time lag. The
cross-correlation gives a  measure of the linear synchronization
between $x$ and $y$. Its absolute value ranges from zero (no synchronization) to one
(maximum synchronization) and it is symmetric: $c_{xy}(\tau) = c_{yx}(\tau)$. 

The sample cross-spectrum is defined as the Fourier transform of the
cross-correlation or, via the Wiener-Khinchin theorem, as: 

\be
C_{xy}(\omega) =({\mathcal F} x) (\omega) \cdot ({\mathcal F} y)^*(\omega) 
\label{eq:crossspectrum}
\ee

\noindent
where $({\mathcal F}x)$ is the Fourier transform of $x$, $\omega$ are
the discrete frequencies ( $-N/2 < \omega < N/2$) and $^*$ means complex conjugation. For
details of the implementation, see sec. \ref{sec-linear}. The cross-spectrum is
a complex number whose normalized amplitude

\be
\Gamma_{xy}(\omega) = \frac{|C_{xy}(\omega)|}{\sqrt{C_{xx}(\omega)} 
 \cdot  \sqrt{C_{yy}(\omega)}} \ ,
\label{eq:coherence}
\ee

\noindent
is called the coherence function and gives a measure of the linear
synchronization between $x$ and $y$ as a function of the frequency $\omega$.
This measure is very useful when synchronization is limited to some
particular frequency band, as it is usually the case in EEG signals
(see \cite{lopes} for a review).

\subsection{Non-linear interdependences}
\label{sec-h}

From time series measured in two systems $\bX$ and $\bY$, let us
reconstruct delay vectors \cite{takens} $\bx_n=(x_n,\ldots,x_{n-
  (m-1)\tau})$ and $\by_n=(y_n,\ldots,y_{n- (m-1)\tau})$, where
$n=1,\ldots N$, $m$ is the embedding dimension and $\tau$ denotes the
time lag.  Let $r_{n,j}$ and $s_{n,j}$, $j=1,\ldots,k$, denote the
time indices of the $k$ nearest neighbors of $\bx_n$ and $\by_n$,
respectively.

\begin{figure}
\begin{center}
\epsfig{file=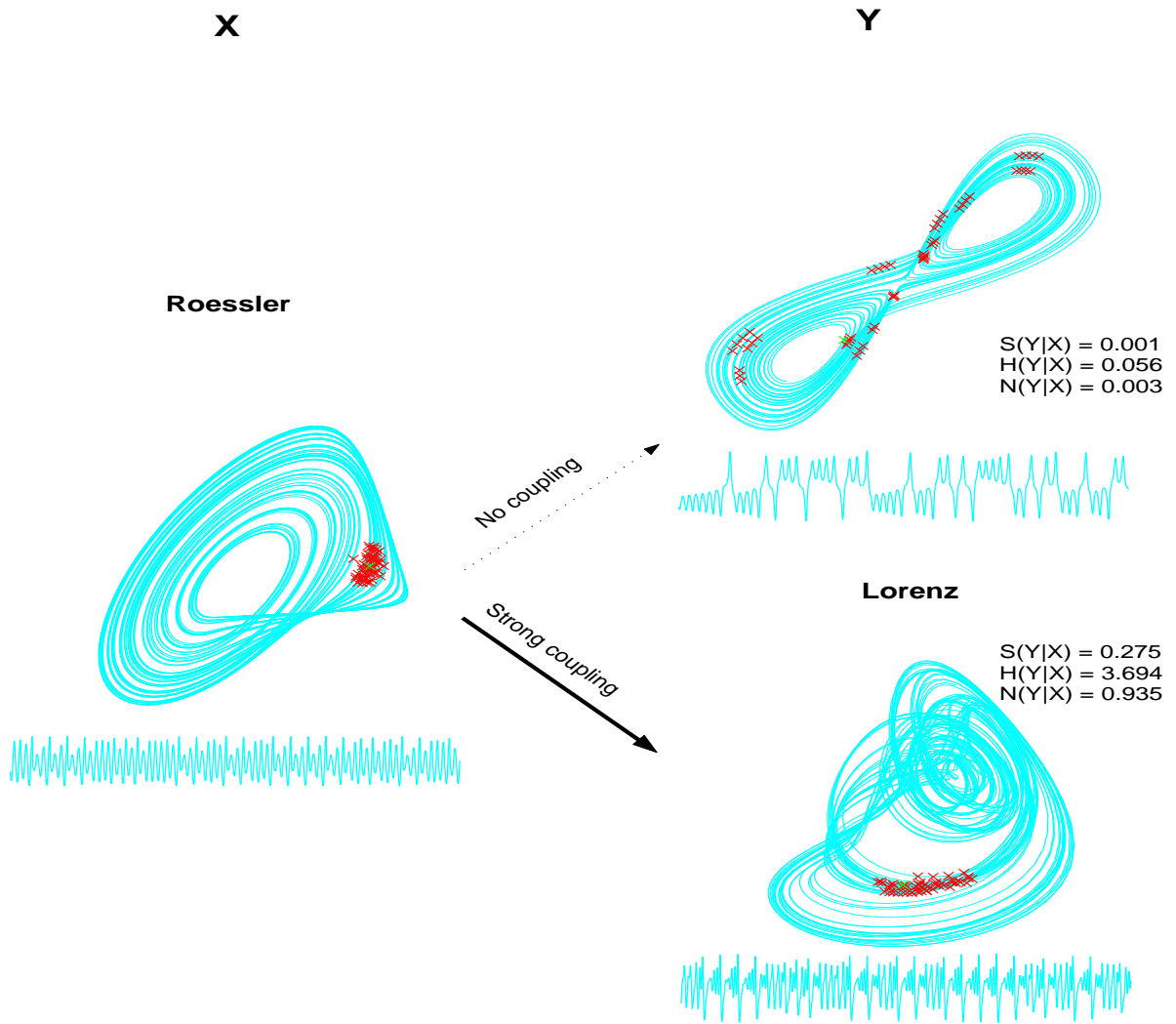, height=14cm,width=16cm,angle=0}
\end{center}
\caption{Basic idea of the non-linear interdependence measures. The
  size of the neighborhood in one of the systems, say $X$, is compared
  with the size of its mapping in the other system. The example shows
  a Lorenz system driven by a Roessler with zero coupling (upper case)
  and with strong coupling (lower case). Below each attractor, the
  corresponding time series is shown. The $(X|Y)$ interdependences are
  calculated in the same way, starting with a neighborhood in $Y$. For
  details see ref. \cite{arnhold,quian}.}
\label{fig:rolo}
\end{figure}

For each $\bx_n$, the squared mean Euclidean distance to its $k$ 
neighbors is defined as

\be
   R_n^{(k)}(\bX)=\frac{1}{k}\sum_{j=1}^{k}{\left( \bx_n - \bx_{r_{n,j}} \right)^2} 
\label{R1}
\ee
and the {\it {\bY}-conditioned} squared mean Euclidean distance is defined by 
replacing the nearest neighbors by the equal time partners of the closest 
neighbors of $\by_n$ see fig. \ref{fig:rolo},

\begin{equation}
   R_n^{(k)}(\bX|\bY)=\frac{1}{k} \sum_{j=1}^{k}{\left( \bx_n - 
          \bx_{s_{n,j}} \right)^2}.                                  
\label{R2}
\end{equation}

If the point cloud $\{\bx_n\}$ has an average squared radius $R(\bX) 
= \frac{1}{N} \sum_{n=1}^N R^{(N-1)}_n (\bX)$, 
then $ R_n^{(k)}(\bX|\bY) \approx R_n^{(k)}(\bX) \ll R(\bX)$ 
if the systems are strongly correlated, while 
$R_n^{(k)}(\bX|\bY) \approx R(\bX) \gg R^{(k)} (\bX)$ if they are independent.
Accordingly, we can define an interdependence measure
$S^{(k)}(\bX|\bY)$ \cite{arnhold} as 
\be
   S^{(k)}(\bX | \bY) = \frac{1}{N} 
           \sum_{n=1}^N \frac{R_n^{(k)}(\bX)}{R_n^{(k)}(\bX|\bY)} .
\label{SXY}
\ee
Since $R_n^{(k)}(\bX|\bY)\ge R_n^{(k)}(\bX)$ by construction, we have 
\be
   0 < S^{(k)}(\bX | \bY) \le 1.
\ee
Low values of $S^{(k)}(\bX | \bY)$ indicate independence between $\bX$
and $\bY$, while high values indicate 
synchronization (reaching maximum when $S^{(k)}(\bX|\bY)\to 1$).

Following ref.\cite{arnhold,quian} we define another 
non-linear interdependence measure $H^{(k)}(\bX|\bY)$ as
\be
      H^{(k)}(\bX | \bY) = \frac{1}{N} \sum_{n=1}^N 
        \log \frac{R_n(\bX)}{R_n^{(k)}(\bX|\bY)}
                                                    \label{HXY}
\ee
This is zero if $\bX$ and $\bY$ are completely independent, while it
is positive if nearness in $\bY$ implies also nearness in $\bX$ for
equal time partners.  It would be negative if close pairs in $\bY$
would correspond mainly to distant pairs in $\bX$. This is very
unlikely but not impossible. Therefore, $H^{(k)}(\bX | \bY)=0$
suggests that $\bX$ and $\bY$ are independent, but does not prove it.
This is one main difference between $H^{(k)}(\bX | \bY)$ and the mutual
 information, to be defined in sec. \ref{sec-mi}. The latter is strictly positive whenever $\bX$ and $\bY$
 are not completely independent. As a consequence, mutual information
 is quadratic in the correlation $P(\bX,\bY)-P(\bX)P(\bY)$ for weak
 correlations ($P$ are here probability distributions), while
 $H^{(k)}(\bX | \bY)$ is linear. Thus $H^{(k)}(\bX | \bY)$ is more
 sensitive to weak dependences which might make it useful in
 applications. Also, it should be easier to estimate than mutual
 informations which are notoriously hard to estimate reliably as we will see later.
  
In a previous study with coupled chaotic systems \cite{quian}, $H$ was
more robust against noise and easier to interpret than $S$, but with
the drawback that it is not normalized. Therefore, we propose a new
measure $N(\bX|\bY)$ using also a different way of averaging,  

\be 
N^{(k)}(\bX | \bY) = \frac{1}{N} \sum_{n=1}^N
\frac{R_n(\bX) - R_n^{(k)}(\bX|\bY)}{R_n(\bX)} \ \,
\label{NXY}
\ee

\noindent
which is normalized (but as in the case of $H$, it can be slightly
negative) and in principle more robust than $S$. 

The opposite interdependences $S(\bY|\bX)$, $H(\bY | \bX)$ and $N(\bY
| \bX)$ are defined in complete analogy and they are in general not
equal to $S(\bX|\bY)$, $H(\bX | \bY)$ and $N(\bY | \bX)$,
respectively.  The asymmetry of $S$, $H$ and $N$ is the main
advantage over other non-linear measures such as the mutual
information and the phase synchronizations defined in sections
\ref{sec-ps} and \ref{sec-wps}. This asymmetry can give information
about driver-response relationships \cite{arnhold,quian,andreas}, but
can also reflect the different dynamical properties of each data
\cite{arnhold,quian}. To address this point we will compare
results with synchronization values obtained from time shifted signals
used as surrogates.

Figure \ref{fig:rolo} illustrates the idea of how the non-linear
interdependence measures work. Let us consider a Lorenz and a Roessler
system that are independent (upper case, no coupling) and a second
case with the Roessler driving the Lorenz via a strong coupling (lower
plot). For a detailed study of synchronization between these systems
refer to \cite{quian}. Given a
neighborhood in one of the attractors, we see how this neighborhood
maps in the other. If the point cloud is still a small neighborhood
(lower plot), the systems are synchronized. On the other hand, if the
points are spread over the attractor (upper plot), the systems are
independent. The three measures described  $S$, $H$ and
$N$, are just different ways of normalizing these ratio of distances.

\subsection{Mutual Information}
\label{sec-mi}

The previous measures of synchronization were based on similarities in
the time and frequency domain (sec. \ref{sec-lin}) or on similarities
in a phase space (sec. \ref{sec-h}). In this section we describe an
approach to measure synchronization by means of information-theoretic
concepts. Let us suppose we have a discrete random variable $X$ with
$M$ possible outcomes $X_1,\ldots,X_M$, obtained e.g. by a partition
of $X$ into $M$ bins. Each outcome has a probability $p_i,
i=1,\ldots, M$, with $p_i \geq 0 ~\forall i$ and $\sum p_i = 1$.  A
first estimate is to consider $p_i = n_i/N$, where
$n_i$ is the number of occurrences of $X_i$ after $N$ samples.  From
this set of probabilities the Shannon entropy is defined as:

\be
I(X) = - \sum_{i=1}^M p_i ~ \log~ p_i
\label{eq:shannon}
\ee

\noindent
The Shannon entropy is positive and measures the information content
of $X$, in bits, if the logarithm is taken with base 2. When finite
samples $N$ are considered, the naive definition $p_i=n_i/N$ may not
be appropriate. Grassberger \cite{grass1} introduced a series of
correction terms, which are asymptotic in $1/N$. The first and most
important term 
essentially gives

\be
I(X) \approx \sum_i \frac{n_i}{N} ( \log~N - \Psi(n_i) ) \ \,
\label{eq:correction}
\ee

\noindent
with $\Psi(x) = d \log \Gamma (x) / dx \approx \log x - 1/2 x$ for large
$x$. 

Let us now suppose we have a second discrete random variable $Y$,
whose degree of synchronization with $X$ we want to measure.
The joint entropy is defined as:

\be
I(X,Y) = - \sum_{i,j} p_{ij}^{XY} ~ \log~ p_{ij}^{XY}
\label{eq:jshannon}
\ee

\noindent
where $p_{ij}^{XY}$ is the joint probability of $X=X_i$ and
$Y=Y_j$. If the systems are independent we have $p_{ij}^{XY} =
p_i^{X} \cdot p_j^{Y}$ and then, $I(X,Y) = I(X) + I(Y)$.
Thus, the mutual information between $X$ and $Y$ is defined as

\be
MI(X,Y) = I(X) + I(Y) - I(X,Y)  \ \,
\label{eq:mi}
\ee

\noindent
which indicates the amount of information of $X$ we obtain by
knowing $Y$ and vice versa. If $X$ and $Y$ are independent,
$MI(X,Y) = 0$ and otherwise, it will take positive values with a
maximum of $MI(X,X) = I(X)$ for identical signals. Note also
that $MI$ is symmetric, i.e. $MI(X,Y)=MI(Y,X)$.  Schreiber
extended the concept of $MI$ and defined a {\it transfer entropy}
\cite{thomas}, which has the main advantage of being asymmetric and
can in principle distinguish driver-response relationships. Another
asymmetric measure based on the $MI$ has been proposed by Palus
\cite{palus}.

Mutual information can also be regarded as a Kullback-Leibler entropy
\cite{gray,kl}, which is an
entropy measure of the similarity between two probability
distributions. To illustrate this, we rewrite eq.(\ref{eq:mi}) in the
form

\be
MI(X,Y) = \sum p_{ij}^{XY} ~ \log  
\frac{p^{XY}_{ij}} {p_i^{X} \cdot p_j^{Y}} 
\label{eq:kl}
\ee

\noindent
Then, considering a probability distribution $q^{XY}_{ij} =
p_i^{X} \cdot p_j^{Y}$ (which is the correct probability
distribution if the systems are independent), eq. (\ref{eq:kl}) is a
Kullback-Leibler entropy and measures the difference between the
probability distributions $p_{ij}^{XY}$ and $q_{ij}^{XY}$
\cite{grass}.  In other words, $MI(X,Y)$ measures how different is
the true joint probability distribution $p_{ij}^{XY}$ from another
in which independence between $X$ and $Y$ is assumed.

We previously mentioned that each $p_i$ can be obtained by a partition
of $X$. In our case, $X$ is the space of time-delay vectors $\bx_n$ as in
section \ref{sec-h}. In
principle, we can calculate  $p_i$ by box counting.
But it was shown in \cite{grass2,schuster} that the Shannon entropies
(eq.~(\ref{eq:shannon})) can be
calculated from the first order correlation integral $C^1(X,\delta)$,
which gives more accurate results \cite{schuster,grass,theiler}.
Thus, instead of calculating probabilities within boxes of a fixed
grid with sidelength $\delta$, we compute probabilities within
neighborhoods of a certain radius $\delta/2$ around each point
\cite{schuster}. Therefore we have: 

\be
I(X;\delta) = - \frac{1}{N} \sum_{i=1}^{N} \log p_i
\label{corr}
\ee

\noindent
with $p_i \simeq \frac{n_i}{N}, n_i = \sum_j \Theta (\delta /2 
- \| \bx_i - \bx_j \| )$ and $N$ the number of embedding vectors. In this case, we can also
introduce finite sample corrections which give \cite{grass1}

\be
I(X; \delta) = - \frac{1}{N} \sum_{i=1}^N ( \Psi (n_i +1) - \log N )
\label{eq:correctedi}
\ee

\subsection{Phase synchronization from the Hilbert Transform}
\label{sec-ps}

Given a univariate measurement $x(t)$ (with continuous t) we
first define the analytic signal $Z_x(t) =
x(t) + i ~\tilde{x}(t) = A_x^H(t) e^{i \phi_{x}^H(t)}$, where
$\tilde{x}(t)$ is the Hilbert Transform of
$x(t)$ \cite{rosemblum},  

\be
\tilde{x}(t) \equiv ({\mathcal H} x)(t) = \frac{1}{\pi} P.V. \int_{-\infty}^{+\infty}
\frac{x(t')}{t-t'} \ dt' \ \ ,
\label{eq:hilbert}
\ee

\noindent
(P.V. means Cauchy principal value). Analogously, we define $A_y^H$ and
$\phi_y^H$ from $y(t)$\footnote{In the actual implementation, where
  $x(t)$ and $y(t)$ are only known at discrete times, we calculate
  $\tilde{x}_n$ from the Fourier transform, as described in \cite{rosemblum}.}.
We say that the $x$ and $y$ are $n:m$ synchronized, if the $(n,m)$
 phase difference of their
analytic signals, $\phi_{xy}^H(t) \equiv n \phi_x^H(t) -
m \phi_y^H(t)$,
with $n, m$ some integers,  remains bounded for all $t$. 
Thus, we define
a phase synchronization index as \cite{pikovsky,florian}

\be
\gamma_{\rm H} \equiv | \langle e^{i \phi_{xy}^H (t)} \rangle_t| =  \sqrt{ \langle \cos{\phi_{xy}^H (t)} \rangle_t^2 +  
\langle \sin{\phi_{xy}^H (t)} \rangle_t^2 }
\label{eq:ps}
\ee

\noindent
(brackets denote average over time). By construction, $\gamma_H$ will
be zero if the phases are not synchronized at all and will be one when
the phase difference is constant (perfect synchronization).  The key
feature of $\gamma_H$ is that it is only sensitive to phases,
irrespective of the amplitude of each signal. This feature has been
illustrated in \cite{rosemblum} and following papers (see
\cite{pikovsky}) with bidirectionally coupled R\"ossler systems.
Another important feature of $\gamma_H$ is that it is parameter free.
However, if the signals to be analyzed have a broadband or a
multimodal spectrum, then the definition of the phase can be
troublesome and pre-filtering of the signals might be necessary. Of
course, it should be checked that the filter to be used does not introduce
phase distortions.

Tass and coworkers \cite{tass} defined another phase synchronization
measure from the Shannon entropy of the distribution of
$\phi_{xy}^H(t)$. The range of $\phi'=\phi_{xy}^H (mod,2\pi)$ is first
divided into $M$ bins. Let $p_k$ be the probability that $\phi'$ is in
the bin $k$ at any random time. Then,       

\be
\gamma_{{\rm H-Sh}} = \frac{S_{{\rm max}}-S}{S_{{\rm max}}} \ , \ \ \ \ S=-\sum_{k=1}^M p_k \cdot
\ln{p_k}
\label{eq:shannonps}
\ee

\noindent
and $S_{max}=\ln{M}$. It ranges from zero for an
uniform distribution of $\phi_{xy}^H$, to one if the distribution is a
delta function. The advantage over $\gamma_{\rm H}$ is that
$\gamma_{\rm H}$ can
underestimate phase synchronizations when the distribution of
$\phi_{xy}^H$ has more than one peak. This corresponds to the case where
the phase difference remains fairly stable but occasionally ``jumps"
between different values\cite{zaks}.  
Although the signals are synchronized (except at
the times of the jumps), the phases $\phi_{xy}^H(t)$ can
cancel in the time average of eq.(\ref{eq:ps}), thus giving a low
$\gamma_{\rm H}$  \footnote{
A multimodal distribution of the phases can also appear if we
look e.g. for a $1:1$ synchronization but the real relationship is
 $1:2$.}.  
We also calculated another quantification proposed in
\cite{tass} defined from conditional probabilities between
$\phi_x^H(t)$ and $\phi_y^H(t)$, but results were very similar to those
obtained with $\gamma_{\rm H}$ and will be not further reported.

\subsection{Phase synchronization from the Wavelet Transform}
\label{sec-wps}

Another phase synchronization measure defined from the Wavelet
Transform ($\gamma_{\rm W}$) has been recently introduced by Lachaux et
al.\cite{lachaux,lachaux1}. It is very similar to $\gamma_{\rm H}$,
 with the only difference that the
phases are calculated by convolving each signal with a complex wavelet
function $\Psi(t)$ \cite{grossman1}

\be
\Psi(t) = (e^{i \omega_0 t} - e^{-\omega_0^2
    \sigma^2 / 2 }) \cdot e^{-t^2 / 2 \sigma^2 } \ \ , 
\label{eq:morlet}
\ee

\noindent
where $w_0$ is the center frequency of the wavelet and $\sigma$
determines its rate of decay (and by the uncertainty principle, its
frequency span)\footnote{Instead of eq.(\ref{eq:morlet}), the authors of \cite{lachaux,lachaux1}
   used a Morlet wavelet i.e. $\Psi(t)
  = e^{i\omega_0 t} \cdot e^{-t^2 / 2 \sigma^2}$, which satisfies the
  zero mean admissibility condition of a wavelet only for large
  $\sigma$. Since in our case we will use a low $\sigma$ (i.e.  a
  $\Psi$ with few significant oscillations, see sect.\ref{sec-wps}),
  an additional negative term is introduced. When $\sigma$ is small,
  disregarding this term can introduce spurious effects, especially if
  the signal to be analyzed has non-zero mean or low frequency
  components. We do not need a normalization term in
  eq. (\ref{eq:morlet}) because we will be interested only in phases. }.  


The convolution of $x(t)$ and $y(t)$ with $\Psi(t)$ gives 
two complex time series of
wavelet coefficients 

\be
W_x(t) = ( \Psi \circ x )_{(t)} =
\int \Psi(t')~ x(t'-t) \ dt' = A_x^W(t) \cdot  e^{i\phi_x^W(t)} \ ,
\label{eq:wt}
\ee

\noindent
($W_y(t)$ is defined in the same way from $y(t)$) from which we can again
calculate the phase differences $\phi_{xy}^W(t) \equiv \phi_x^W (t) - \phi_y^W
(t)$ and define a  phase synchronization factor ($\gamma_{\rm W}$) as in eq.
(\ref{eq:ps}), or from the Shannon entropy of the distribution of
$\phi_{xy}^W(t)$ ($\gamma_{\rm W-Sh}$) as in eq.(\ref{eq:shannonps}). 

The main difference with the measures defined by using the Hilbert
transform is that a central frequency $\omega_0$ and a width 
$\sigma$ for the wavelet function should be chosen, and therefore
$\gamma_{\rm W}$ and $\gamma_{\rm W-Sh}$ will be sensitive only to
phase synchronizations in a certain frequency band.  
In particular,
DeShazer et. al. \cite{ott} recently analyzed phase synchronization in
coupled laser systems
defining the phases both from a Gabor (similar to eq.(\ref{eq:morlet})) and a
Hilbert transform.  In the first case
they distinguished a phase synchronization at 140 Hz, something not
seen when using the Hilbert transform.  The difference between both
approaches, of course, does not imply that one measure is superior to
the other. There are cases in which one would like to restrict the
analysis to a certain frequency band and other cases in which one
would prefer to have a method that is parameter free, as $\gamma_{\rm
  H}$. In fact, in section \ref{sec-relation} we will show that there
is a close relation between both methods.

\subsection{Relation between the phase synchronization measures}
\label{sec-relation}

In sections \ref{sec-ps} we already mentioned that in some cases it
might be necessary to pre-filter the signals before applying the
Hilbert Transform, while for the Wavelet Transform a center frequency
(and frequency width) should be chosen beforehand. 
In fact, the phases defined by the complex wavelet transform
$\phi_x^W$ and by the Hilbert transform $\phi_x^H$ are closely
related. Indeed, the real part of $W_x(t)$ can be considered as a
band-pass filtered signal. From it, we can form the Hilbert transform 

\be
\tilde{W}_x(t) = ( {\mathcal H} ~ {\rm Re[W_x]} ~)(t) \ \ , 
\ee
and a phase by
\be
{\rm Re[W_x]}(t) + i ~ \tilde{W}_x(t) = {A_{\rm Re[W_x]}^H}(t) \cdot
e^{i{\phi_{\rm Re[W_x]}^H}(t)} . 
\ee

\noindent
Let us now recall the definition of analytic signals. A complex function $g(t)$
is an analytic signal if it satisfies $({\mathcal F} g) (\omega) = 0~~
\forall~ \omega < 0 $  \cite{cohen}. If $g$ is analytic, then  
${\rm Im} [g(t)] = \tilde{g}(t) \equiv ( {\mathcal H} ~ {\rm Re} [g]) (t)$. 
If a wavelet function $\Psi$ is analytic, then  
$W_x (t) = ( \Psi \circ x )_{(t)}$ is also analytic\footnote{
Taking the Fourier Transform we get
$({\mathcal F}~ W_x) (\omega) = ({\mathcal F} ~( \Psi \circ x )_{(t)} )(\omega) =
({\mathcal F} \Psi)(\omega) \cdot ({\mathcal F} x)(\omega) = 0 ~~\forall ~\omega <0$, 
where we used the Fourier convolution theorem and that $\Psi$ is
analytic.}. 
In this case $\tilde{W}_x(t) \equiv {\rm Im} [W_x(t)]$ and
$\phi_{\rm Re[W_x]}^H(t) \equiv \phi_x^W(t)$, as defined in eq.(\ref{eq:wt}).
Since the corrected Morlet wavelet of eq.(\ref{eq:morlet}) is
approximately analytic\footnote{
The Morlet wavelet  tends to the
analytic signal for large $\omega_0$ and low $\sigma$ \cite{cohen}.}
we have $\phi_{\rm Re[W_x]}^H(t) \cong \phi_x^W(t)$ to very good approximation.
Since as we mentioned, $W_x(t)$ acts as a band pass filter of $x(t)$,
then $\phi_{x}^H(t) \cong \phi_x^W(t)$
 as long as for the first one  the signal is pre-filtered with the same wavelet function
used for calculating the latter.

It is important to remark that the previous result is not limited to
complex Morlet wavelets and can be extended to other wavelet 
functions. In particular, from a real wavelet function $\Psi(t)$ we
can construct an analytic signal by using the
Hilbert transform, i.e. $\Psi'(t) \equiv \Psi (t) + i~({\mathcal H}
\Psi)(t)$, which satisfies that $W_x(t) = ( \Psi' \circ x ) (t)$ is analytic.  
Then, from $W_x(t)$ we can define a phase and e.g. study the phase
synchronization with another signal $y(t)$. The important advantage is
that we have the freedom of defining the phase from a particular
wavelet function, chosen from a dictionary of available
wavelets according to the signal to be studied.  This can be
interesting in cases in which defining a phase from the Hilbert
transform is troublesome or if conventional filters are not well suited.

\section{Details of the data}
\label{sec-data}

We will analyze the synchronization between two EEG channels in three
different data sets \cite{www}. The EEG signals were obtained from
electrodes placed on the left and right frontal cortex of male adult
WAG/Rij rats (a genetic model for human absence epilepsy)
\cite{giles}. Both signals were referenced to an electrode placed at
the cerebellum, they were filtered between 1-100 Hz and digitized at
200 Hz.

In a previous study \cite{sleepwake}, the main objective of this set up
was to study changes in synchronization after unilateral lesions with
ibothenic acid in the rostral pole of the reticular
thalamic nucleus. To achieve this, synchronization
was first assessed visually by looking for the simultaneous
appearance of spike discharges\footnote{More properly, 
  ``spike-wave discharges" but for simplicity we will call them spikes
  in the remaining of the paper.} and then it was further quantified by
calculating both a linear cross-correlation and the non-linear
interdependence measure $H$ defined in the previous section. For the
quantitative analysis, for each rat and condition, 10 data segments
pre- and 10 segments post-lesion were analyzed, five of these segments
corresponding to normal EEGs and the other five containing spike
discharges. The length of each data segment was 5 seconds (i.e.
1000 data points), this being the largest length in which the signals
containing spikes could be visually judged as stationary.  In all 7
rats studied, it was found that synchronization significantly
decreased after the lesions in the reticular thalamic nucleus
\cite{sleepwake}.  Moreover, changes shown with the non-linear
synchronization $H$ were more pronounced than those found with the
cross-correlation.  In the following section we will analyze in detail
three of these EEG segments.

\section{Synchronization in the EEG data}
\label{sec-results}
 
In Fig.\ref{fig:examples} we show the right and left channels of three
of the (pre-lesion) EEG signals described in the previous section.
The first case (example A) corresponds to a normal EEG, and in the
remaining two cases the signals have spike discharges (examples B and
C).  Spikes usually appear due to a local synchronization of neurons
in the neighborhood of the electrode at which they are recorded. Since
epilepsy is related to an abnormal synchronization in the brain,
spikes are usually considered as a landmark of epileptic activity.  A localized appearance
of spikes can delimit a zone with abnormal discharges (but this will
not necessarily be the epileptic focus). On the contrary, if spikes
are observed over the whole set of electrodes, abnormal
synchronization is said to be global. This concept seems to be
obvious, but it has some subtleties as we will see in the following.
Let us analyze examples B and C. In both cases we see spikes at the
left and right electrodes. As we said, this will point towards a
global synchronization behavior. However, a more detailed analysis
shows that the spikes of example B are well synchronized and in
example C they are not. Indeed, in example C the spikes have slightly
different time lags between the right and left channels. This is of
course not easily seen in a first sight. For making clear this point,
we picked up the spikes of examples B and C and we noted the times of
their maximum for the right and left channels. We then calculated the
lag between the spikes in the two channels and its standard deviation
with time. For the case B, the lag was very small and stable, mainly
between -5 to 5 ms (i.e. of the order of the sampling rate) and the standard
deviation was of 4.7 ms. For the case C, the lag was much more
unstable and covered a larger range (between -20 to 50ms). In this
last case the standard deviation was of 14.9 ms. This shows that in
example B the simultaneous appearance of spikes is correlated with a
global synchronization, while in example C bilateral spikes are not
synchronized (i.e. we have local synchronization for both channels,
but no global synchronization).  In the case of example A, due to its
random-like appearance it is difficult to estimate the
level of synchronization by visual inspection. However, we can already observe some
patterns appearing simultaneously in both the left and right channels,
thus showing some degree of interdependence.

Summarizing, we may say that example B seems the most ``ordered"
and synchronized. Among the other two examples, A {\it looks} definitely
more disordered than C, but a closer look raises doubts and a formal
analysis is asked for.

\subsection{Linear measures}
\label{sec-linear}

\begin{figure}
\begin{center}
\epsfig{file=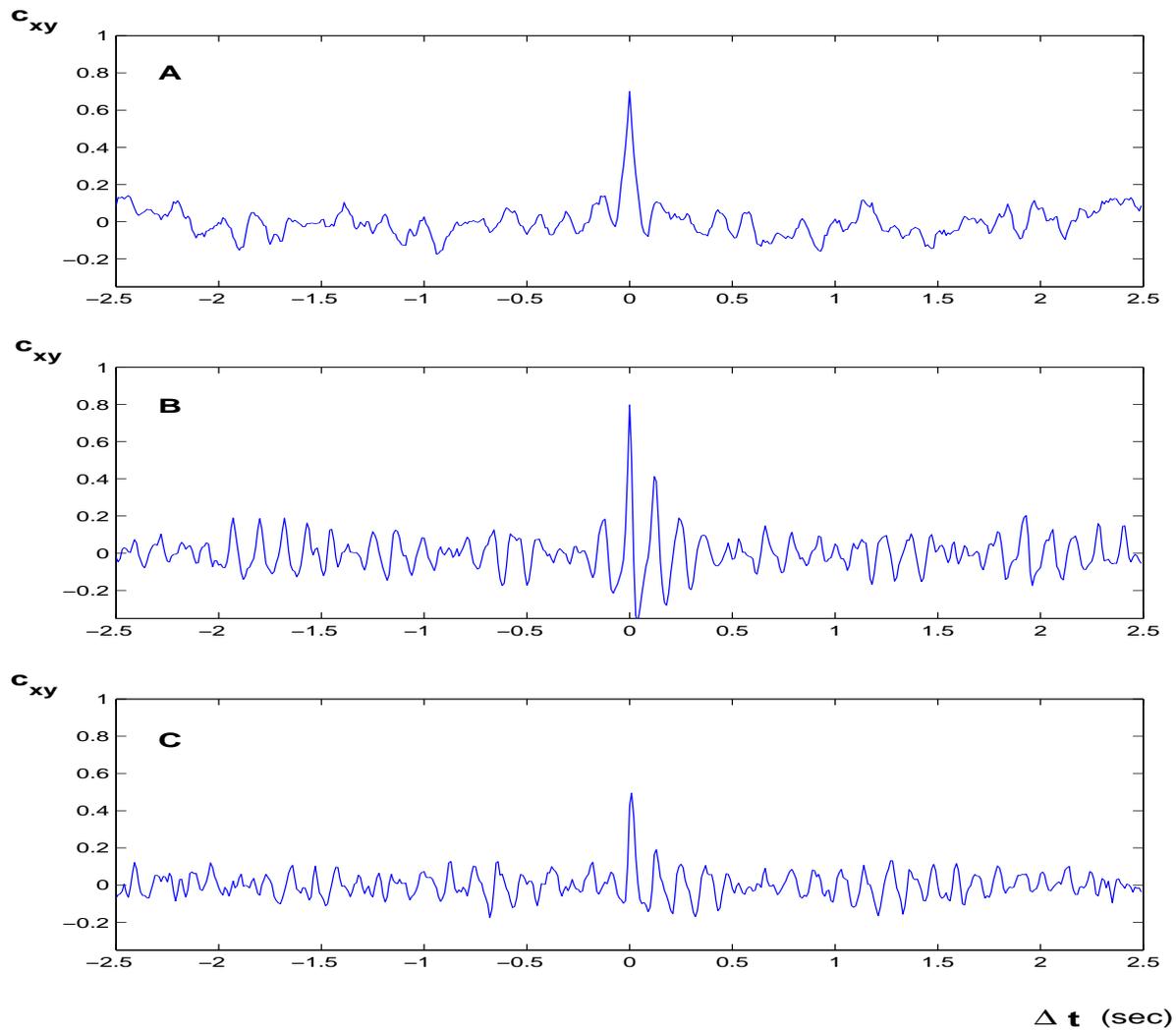, height=14cm,width=16cm,angle=0}
\end{center}
\caption{Cross-correlations
  between the right channel and shifted versions of the left
  one. Note that the difference between the three signals are of the order of fluctuations when shifting.}
\label{fig:shift_cross}
\end{figure}

The second column of Table \ref{tab:h} shows the zero lag
cross-correlation values for the three examples. As stated in
eq.(\ref{eq:cross}), the calculation of the cross-correlation requires
a normalization of the data.  We note that the tendency is in
agreement with what we expect from the arguments of the previous
section (i.e. $B > A > C$). However, the difference between cases A
and B is relatively small. To get more insight, in
Fig.\ref{fig:shift_cross} we plot the cross-correlation as a function of
time shifts between the two channels. For the shifted versions, we
used periodic boundary conditions. For large enough shifts, the
synchronization will in principle be lost and the values obtained will
give an estimation of the zero synchronization level, which we will
call background level,  and its
fluctuation (i.e. we use the shifted versions as surrogates). We
observe that the synchronization drops to a background level for
shifts larger than 50 data points (i.e. 250ms). The average of this
background level is zero, but the fluctuations are quite large. Taking
these fluctuations as an estimation of the error, we see that 
cross-correlation does not distinguish between cases A and B.

We also note that the cross-correlation shows oscillations when
shifting, most clearly in case B. These oscillations have the same
period of the spikes and might put into doubt the idea of
considering the shifted signals as surrogates.  We therefore
re-calculated the cross-correlation but taking the left channel
signals from other data segments of the same rat (for each rat we
had 5 segments with spikes and 5 of normal EEG before the lesions in
the thalamus) and corresponding to the same condition (pre-lesion,
normal EEG for example A and EEG with spikes for examples B and C). In
all cases, the background level and its fluctuations were of the order
of those shown in fig.\ref{fig:shift_cross}.  This indicates that shifted
signals can be used as surrogates in spite of the oscillations.

\begin{figure}
\begin{center}
\epsfig{file=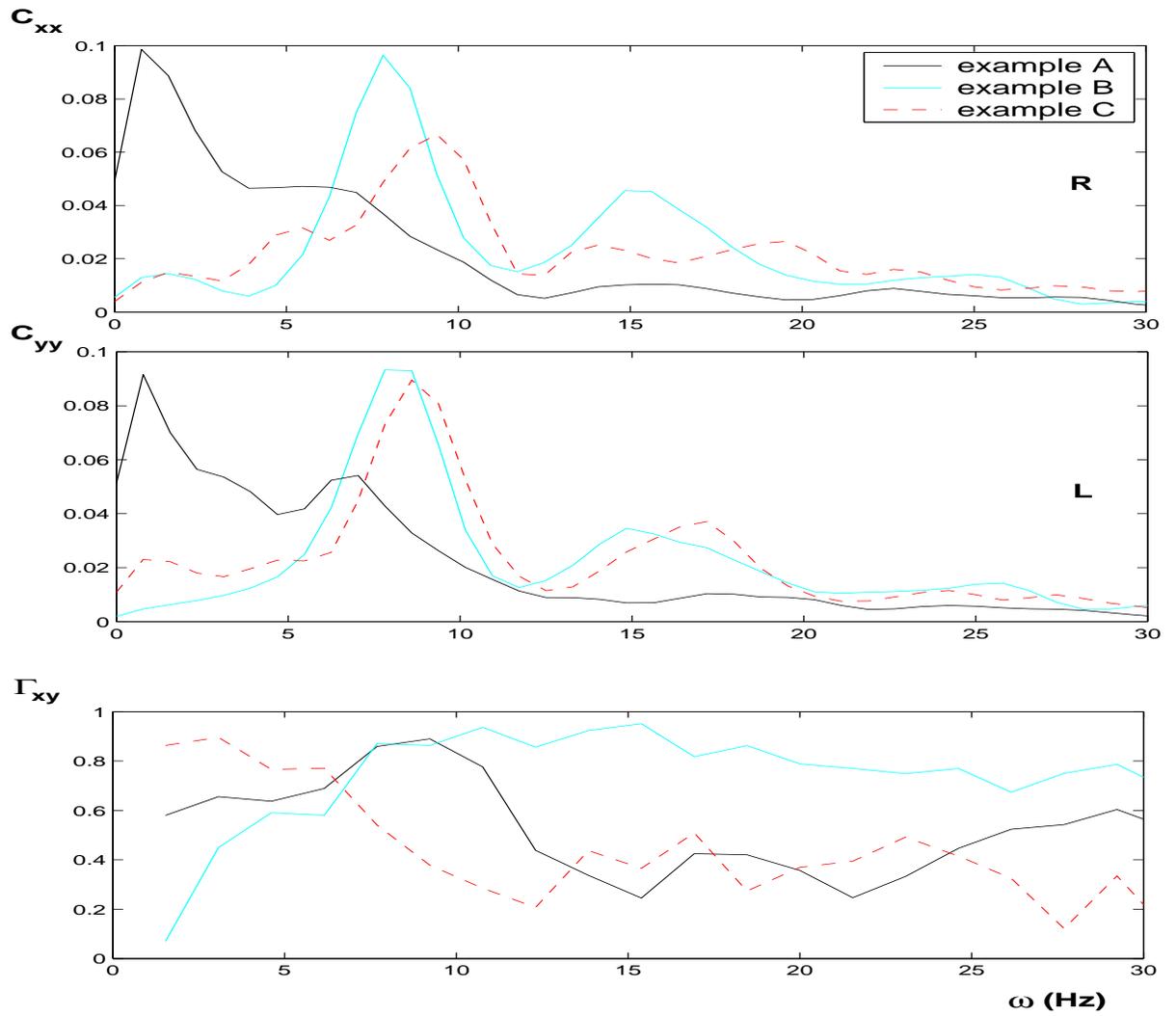, height=14cm,width=16cm,angle=0}
\end{center}
\caption{Power spectra of the right and left channels ($C_{xx}$ and
  $C_{yy}$; upper plots) and the
  corresponding coherence function ($\Gamma_{xy}$; lower plot).}
\label{fig:fourier}
\end{figure}

Figure \ref{fig:fourier} shows the spectral estimates for the three
examples. The two upper plots correspond to the power
spectra of the right and left channels and the lower plot to the
corresponding coherence function (\ref{eq:coherence}). Each spectrum
($C_{xx}$, $C_{yy}$ and $C_{xy}$) was estimated using the Welch
technique\footnote{without this segmentation technique, the coherence
  function (eq.(\ref{eq:coherence})) would be always equal to one.},
i.e. the data is divided into $M$ segments and then
$C_{xx}=\sum_{i=1}^M C_{xx,i}$.  We used half overlapped segments of
128 data points tapered with a Hamming window.  Example A has both in
the right and left channels a power spectrum resembling a power law
distribution, with its main activity concentrated between $1-10$Hz.
The coherence function shows a significant interaction for this range
of frequencies.  Examples B and C show a more localized distribution
in the power spectrum. In both examples and for both channels there is
a peak between $7-10$Hz and a harmonic at about $15$Hz.  In
agreement with previously reported results \cite{drinkenburg}, these peaks
correspond to the spikes observed in Fig.\ref{fig:examples}. We can
already see from the power spectra that the matching between right and
left channels in example B is much clearer than in example C. This is
correlated with the larger coherence values of example B, showing a
significant synchronization for almost the whole frequency range.
On the other hand, the coherence is much lower for example C and it
seems to be significant only for low frequencies (up to 6Hz). As in
the case of the cross-correlation, the coherence function for $\omega
\leq 11 {\rm Hz}$ does not distinguish well between examples A and B.
There is only a difference for frequencies larger than 11Hz, but this
just reflects the lack of activity in this frequency range for example
A, whereas in example B it corresponds to the synchronization between
the high frequency harmonics of the spikes.  In the third column of
Table \ref{tab:h} we summarize the results obtained with the coherence
function. The values shown correspond to a frequency of 9Hz, the main
frequency of the spikes in examples B and C.

\subsection{Non-linear interdependences}
\label{sec-resultsh}

For calculating the non-linear interdependence measures $S, H$ and $N$ 
between left and right electrodes we first reconstruct the state 
spaces of each signal using a time lag $\tau = 2$ and an embedding
dimension $m=10$. We chose this time lag in order to focus on
frequencies lower than 50Hz (i.e. half the Nyquist frequency) and the
choice of the embedding dimension was in order to have the length of the
embedding vectors about the length of the spikes.  We further chose
$k=10$ nearest neighbors and a Theiler correction for temporal correlations
\cite{theiler1} of $T=50$. These parameters were chosen heuristically
in order to maximize the sensitivity to the underlying
synchronizations, but results were robust against changes of them.
Table \ref{tab:h} summarizes the results for the three examples. We
will first discuss results with the non-linear measures $H$ and $N$.
For both measures, example B has the highest
synchronization due to the presence of phase-locked spike discharges
and example C has a much smaller value.  The synchronization of
example A is between these values. Again, it is interesting to remark
that the non-linear interdependence measures show  the random looking
signal of example A to be more synchronized than
the one with spikes of example C but less than the one in B, something
surprising at a first sight, and not clearly following from  the
cross-correlation or the coherence as shown in section
\ref{sec-linear}.

\begin{figure}
\begin{center}
\epsfig{file=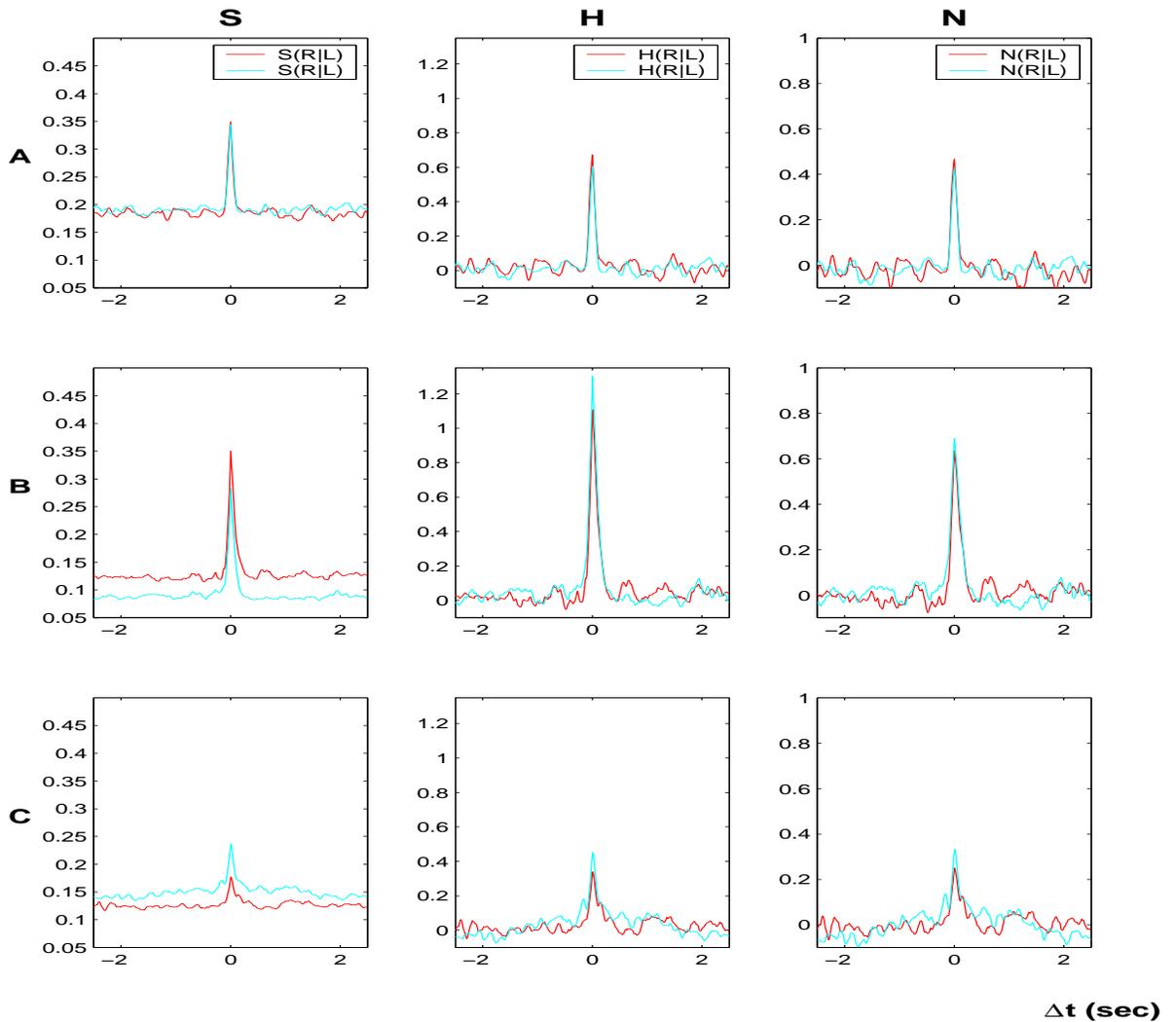, height=14cm,width=16cm,angle=0}
\end{center}
\caption{Non-linear interdependences S, H and N  
  between the right channel and shifted versions of the left
  one. Note that H and N give similar results and can distinguish the three cases. The measure S shows an asymmetry that remains even after shifting. See text for details.}
\label{fig:shift_non}
\end{figure}

As done for the cross-correlation, in Fig.\ref{fig:shift_non} we also
plot the two non-linear synchronizations $H(R|L)$, $N(R|L)$ and
$H(L|R)$, $N(L|R)$ as a
function of time shifts between the two channels. Again, the
synchronization drops to a background level for shifts larger than 50
data points (i.e. 250ms) and the background level is about zero. But in
the case of $H$ and $N$ we observe that the fluctuations are much smaller than
those for the cross-correlation.  In fact, with $H$ and $N$ the
synchronization levels of the three cases are clearly separated, while
the cross-correlation does not distinguish between cases A and B. 
However, even though we expect example B to be the most ordered and
synchronized of all (see sec. \ref{sec-results}), we do not have
objective means for claiming that 
the difference between  examples A and B is significant. So, the
fact that non-linear measures are able to separate the three examples 
might imply a higher sensitivity of these measures in comparison with the linear measures, but it does not prove it.
We also observe some asymmetries in
$H$ and $N$, most pronounced in case C. This might suggest that one of the signals
drives the other (i.e. the focus is on one side).  However, in all
cases this is of the order of the asymmetries seen with the shifted
signals, thus not significant.

The case for the synchronization measure $S$ is quite different. As
seen in Fig.\ref{fig:shift_non}, for examples B and C there is a clear
asymmetry between right and left channels. In contrary to $H$ and $N$, this
asymmetry remains even for large time shifts between the two channels.
Moreover, the background level for the three examples is between $0.1
- 0.2$ and not zero as with $H$.  
Thus, the asymmetries observed in examples B and C reflect
more the individual properties of each channel rather
than a synchronization phenomenon\footnote{As pointed out in
  \cite{arnhold}, precisely such an asymmetry is expected if otherwise
  equal systems are coupled asymmetrically. Thus, if we expect both
  subsystems a priori to have the same complexity, the asymmetry of
  $S$ is a hint to an asymmetric coupling.}. Nevertheless, $H$ and $N$ were clearly
more robust in this respect. 

Again, in order to check for the validity of the shifted signals as
surrogates, we re-calculated $H$, $N$ and $S$ but taking the left channel
signals from other data segments. As in the case of the
cross-correlation, the background level and its fluctuations were of
the order of those shown in fig.\ref{fig:shift_non}.

\subsection{Hilbert phase synchronization}
\label{sec-resultsps}

\begin{figure}
\begin{center}
\epsfig{file=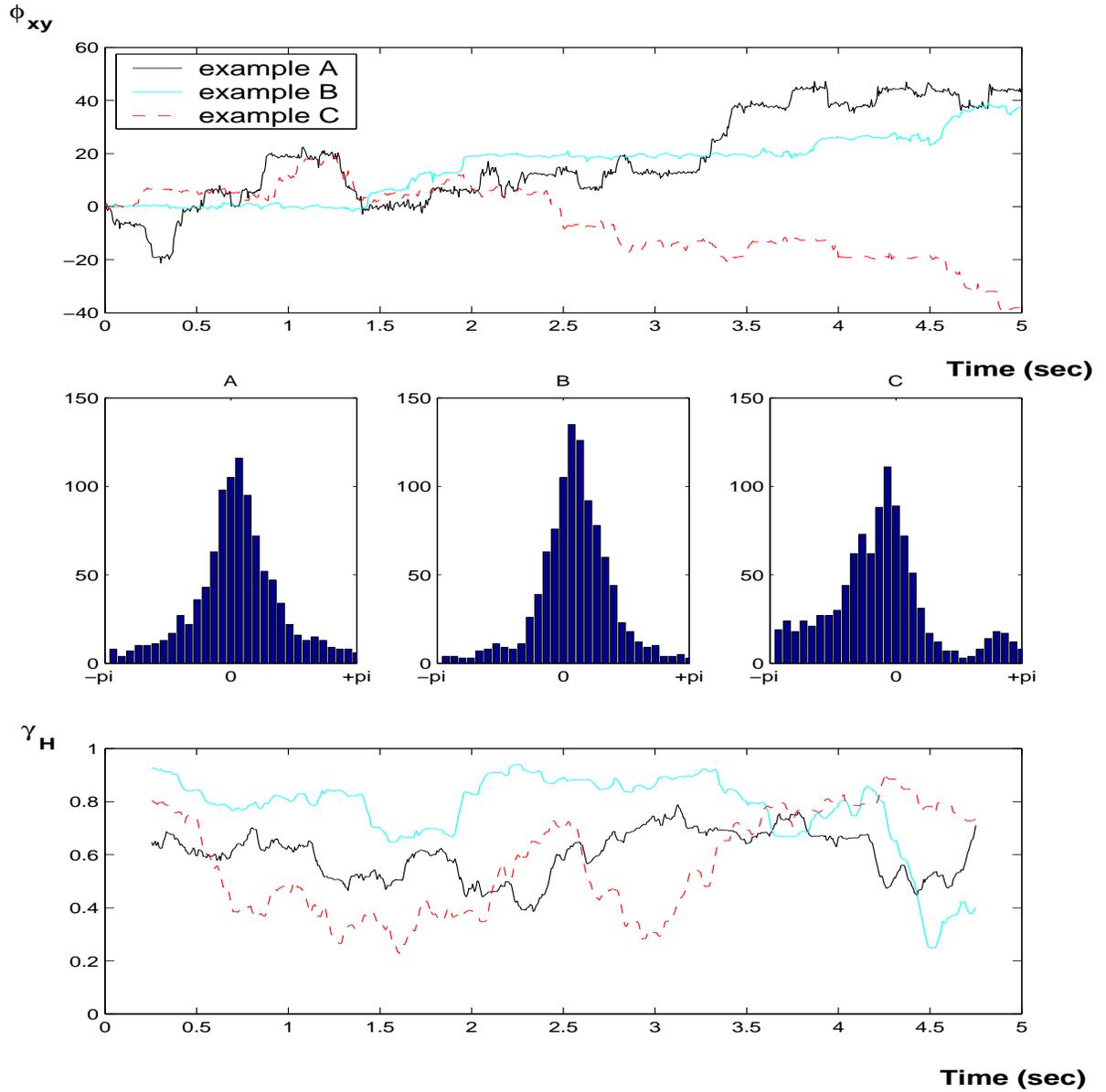, height=16cm,width=16cm,angle=0}
\end{center}
\caption{Time evolution of the (1,1) phase differences, as defined from the Hilbert transform, for the
  three examples of figure \ref{fig:examples} (upper plot), the
  corresponding distributions of the folded phase differences (middle
  plots) and the time evolution of the phase synchronization index
  $\gamma_{\rm H}$ (lower plot).}
\label{fig:phase}
\end{figure}

Prior to the estimation of the phase synchronization measures, each
set of data was de-meaned. No further filtering was applied. Figure \ref{fig:phase}
shows the time evolution of the phases (upper plot) and their
distribution (middle plots) for the three examples. From the
time evolution of the phases we can already see that the phase of
example B is clearly more stable than the other two examples (except
in the last half second, as we will detail later).  Examples A and C
are not so easily differentiated, but in the middle plots we see that
the phase distribution of A is more localized than the one of C. The
values of $\gamma_H$, indicated in Table \ref{tab:h}, are in agreement
with these observations and with the general tendency observed with
the other synchronization measures ($B>A>C$).  
The phase synchronization index defined from the Shannon entropy 
 ($\gamma_{\rm H-Sh}$, defined in eq.(\ref{eq:shannonps}))
shows qualitatively similar results (see Table \ref{tab:h}). 

Since by applying the Hilbert Transform we can calculate an
`instantaneous phase' of the signals, we expect to achieve a very good
time resolution with the phase synchronization measures derived from
them. In the lowest plot of Fig. \ref{fig:phase} we show the time
evolution of $\gamma_{\rm H}$ (the plot for $\gamma_{\rm W-Sh}$ was
qualitatively similar). Each point is calculated for a window of 100
data points. In the first 3 seconds we observe relatively stable
synchronization values for cases A and B.  For the example C we observe a
larger variability due to a progressive phase desynchronization with
a phase reentrainment at about second 2.5. For all the examples, 
synchronization levels oscillate around the average values noted in Table
\ref{tab:h}. After the third second the situation changes. 
Example C becomes more synchronized than the
other two examples and  example B gets more
desynchronized in the last half a second. This is in
agreement with what we see in the original signals in
Fig.\ref{fig:examples}, where it would have been hard to discern at a
first sight by visual inspection.  The possibility to follow phase
synchronization in time is in fact one advantage over
the non-linear interdependences, where a large number of data points
is required for reasonably stable results.

\subsection{Wavelet phase synchronization}
\label{sec-resultswps}

In this case, for calculating the phase of each signal we used a
corrected Morlet wavelet (eq.(\ref{eq:morlet})) with $w_0$ between 1
and 30 Hz and $\sigma= n/6\omega_0$, where $n$ is the number of
significant oscillations of the wavelet function at the $1\%$ level.
We tested different values of $n$ but in the following results with
$n=1$ and $n=3$ will be shown. Larger values of $n$ led to a very bad
time resolution as we detail later. We used zero padding border
conditions and varied $\omega_0$ at 1 Hz intervals.

\begin{figure}
\begin{center}
\epsfig{file=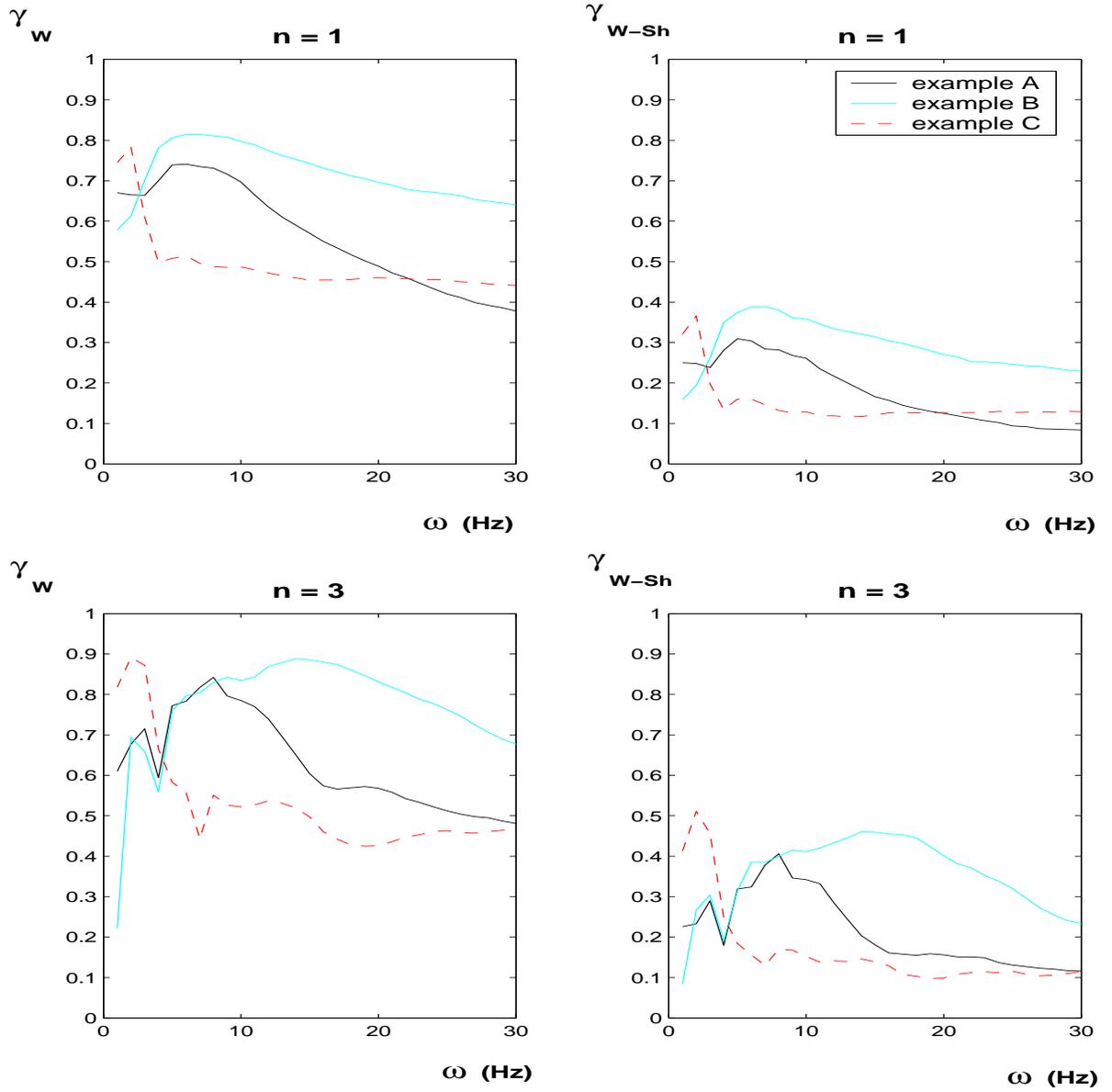, height=16cm,width=16cm,angle=0}
\end{center}
\caption{Phase synchronization indices  $\gamma_{\rm W}$ and
  $\gamma_{\rm W-Sh}$
  defined from the Wavelet Transform for two different wavelet 
  functions ($n=1$ and $n=3$).}
\label{fig:wavelet}
\end{figure}

The phase difference plots (at 10Hz) were indeed very similar to those
shown in Fig.\ref{fig:phase} and will not be discussed further. 
Figure \ref{fig:wavelet} shows the
phase synchronization values $\gamma_{\rm W}$ (left plots) and
$\gamma_{\rm W-Sh}$ (right plots) calculated with a wavelet function
containing 1 significant oscillation ($n=1$; upper plots) and 3
significant oscillations ($n=3$; lower plots).  The values reported in
Table \ref{tab:h} correspond to those obtained with $n=1$ at a
frequency of 10Hz (the frequency of the spikes in examples B and C,
but results are qualitatively the same between $5-15Hz$).
These results are very similar to those obtained with the Hilbert
transform and show the same tendency (i.e. $B > A > C$).  However, we
also note that synchronization values are a bit larger than the ones
of $\gamma_H$ and $\gamma_{\rm H-Sh}$. As already shown in section
\ref{sec-relation}, the difference is due to the frequency band
selectivity of $\gamma_{\rm W}$ and $\gamma_{\rm W-Sh}$. We therefore
expect that a pre-filtering of the signals will increase the
synchronization values calculated by using the Hilbert Transform.

With $n = 1$ the three cases are well differentiated both by
$\gamma_{\rm W-Sh}$ and  $\gamma_{\rm W}$.  With $n=3$ the differences
between the synchronization levels of examples A and B is less clear
 for $\gamma_{\rm W}$ and $\gamma_{\rm W-Sh}$.  This is due to the
decrease in time resolution when increasing the number of significant
oscillations of the mother function. Clearly, for the examples
studied, $n=1$ had the best performance (for $n>3$ results get worse
than for $n=3$).  Notice  the
similarity between the lower plots for $n=3$, i.e. the ones with less
resolution,  with the coherence plots
shown in Fig. \ref{fig:fourier}.  This supports the
usefulness of the phase synchronization measures defined from the
Wavelet Transform in comparison with traditional approaches. Finally,
we should also remark that, as shown in section \ref{sec-relation}, we
are not limited to use Morlet wavelets, but we can rather choose
between several wavelet functions depending on the application.

\subsection{Mutual information}
\label{sec-mir}

Let us finally analyze the results obtained with mutual information
for the three EEG signals. For its calculation we used eq.
(\ref{eq:mi}) with each Shannon entropy calculated by means of the
correlation sum (using maximum norm) and the finite samples correction
of eq.  (\ref{eq:correctedi}).  After each data set was normalized,
for embedding the data we used a time lag $\tau = 2$ and embedding
dimensions ranging from $m=1$ (no embedding) to $m=50$.  We further
used a Theiler correction \cite{theiler1} of 10 data points and for
calculating the correlation sum we varied the radius $\delta$ from 0.01
to 0.5 in steps of 0.01.  In figure \ref{fig:mi} we show the results
for $m=1,2,3,4$, the results for larger $m$ had a similar tendency
(see below).  The difficult point when calculating $MI$ is to have a
good estimation of the joint probabilities $p_{ij}^{XY}$ (see
eq.(\ref{eq:jshannon})). These joint probabilities involve a search of
neighbors in a $2m$-dimensional embedding space, and therefore it is
difficult to find enough neighbors and get a good statistic for large
$m$.  We expect this restriction to be more relevant in the signals
with spikes, due to their inhomogeneous distribution in state space.

In line with the previous argument, due to the small number of data
points we could not get robust estimates of synchronization in the
three examples analyzed.  As seen in Fig. \ref{fig:mi}, the answer to 
the question of which signal
is more and which is less synchronized dramatically depends  on the
choice of $m$ and $\delta$.  We observe the same tendency as with the
previous measures ($B>A>C$) only for $m=1$ and $\delta > 0.15$.
  
All previous analysis done in this paper show clear evidence that
example B has the highest synchronization.
For $m=1$ this is the case for $\delta > 0.05$, for $m=2$ it occurs
for $\delta > 0.2$, for $m=3$ at $\delta > 0.45$ and for $m=4$ it does
not occur for the range of $\delta$ shown.
In fact, there is a crossing between the synchronization values of
examples A and B, that takes place at larger $\delta$ for larger
$m$. This simply reflects the impossibility of finding neighbors in the
$2m$-dimensional state space for small $\delta$ and/or large $m$. As mentioned
before, we expect this effect to be less restrictive for the
homogeneous distribution of example A. This explains why example A
always shows the highest synchronization for small $\delta$. 

\section{Conclusions}
\label{sec-conclusions}

We applied several linear and non-linear measures of synchronization
to three typical EEG signals. Besides mutual information, which was
not robust due to the low number of data points, all these measures gave a similar
tendency in the synchronization levels. A similar analysis would have
been impossible by visual inspection. 
Moreover, in one case with bilateral spikes, synchronization was much
lower than expected at a first sight.  Therefore, we claim
that the quantification of synchronization between different EEG
signals can complement the conventional visual analysis and can even
be of clinical value. In particular, this is very important for the study
of epilepsy \cite{vanqyen1,vanqyen2,arnhold,florian} and for the study of
brain processes involving a synchronous activation of different areas
or structures in the brain.

In the last years, mainly two types of non-linear synchronization
measures were proposed, namely, the ones based on phase relationships
(phase synchronization) and the ones based on non-linear
interdependences (generalized synchronization).  It is interesting to
remark that in our study with real data these measures gave similar
results, despite their different definitions and their sensitivity to
different characteristics of the signals. We also show a close
similarity between phase synchronization measures based on the Hilbert
and on the Wavelet Transform. In the particular case of the last one,
we generalize its definition to different wavelet functions that will
be more or less suitable according to the problem under investigation.

We validated the
results obtained with the new non-linear measures by comparing them
with those obtained with traditional methods.  
All measures ranked the synchronization levels of the three examples
in the same way. 
However the separation between them was more pronounced with
non-linear measures.
Although we do not have objective means for claiming that 
the difference between the synchronization of the signals is
significant, this might suggest a higher sensitivity of non-linear
measures. 

Although these results should not be automatically extended to other
signals and problems, they also support the value of non-linear
synchronization measures in real data analysis.

\newpage
\section*{Acknowledgments}

We are very thankful to Dr. Giles van Luijtelaar and to Joyce Welting
from NICI - University of Nijmegen, 
for the data used in this paper. We are also indebt to Dr. Klaus
Lehnertz, Florian Mormann and Giles van Luijtelaar for useful discussions and comments.
One of us (A.K.) acknowledges support from the US civilian research
development foundation for the independent states of the former Soviet
Union, Award nr: REC-006.

\bibliography{synchro}

\newpage
\begin{thebibliography}{10}



\bibitem{fujisaka}
H. Fujisaka and T. Yamada, Prog. Theor. Phys. {\bf 69}, 32 (1983); {\bf 76}, 582 (1986)

\bibitem{yamada}
T. Yamada and H. Fujisaka, Prog. Theor. Phys. {\bf 70}, 1240 (1984); {\bf 72}, 885 (1985)


\bibitem{pikovskygs}
A. Pikovsky, Z. Physik {\bf B 55}, 149 (1984). 

\bibitem{pecora}
L.M. Pecora and T.L. Carroll, Phys. Rev. Lett. {\bf 64}, 821 (1990).

\bibitem{rulkov}
N.F. Rulkov, M.M. Sushchik, L.S. Tsimring, and H.D.I. Abarbanel.
Phys. Rev. E {\bf 51}, 980 (1995).


\bibitem{singer}
C. Gray, P. Koenig, A. Engel and W. Singer, Nature {\bf 338}, 335 (1989).

\bibitem{niedermeyer}
E. Niedermeyer. Epileptic seizure disorders in {\it Electroencephalography: 
Basic Principles, Clinical Applications and Related Fields}, edited by
E. Niedermeyer and F. H. Lopes da Silva (Baltimore, Williams and
Wilkins 3rd ed., 1993), pp 1097.

\bibitem{schiff}
S.J. Schiff, P.~So, T.~Chang, R.E. Burke and T. Sauer. Phys. Rev. E {\bf 54}, 6708 (1996).

\bibitem{vanqyen1}
M.~Le~Van Quyen, C.~Adam, M.~Baulac, J.~Martinerie, and F.J. Varela, Brain Research {\bf 792}, 24 (1998).

\bibitem{vanqyen2}
M.~Le~Van Quyen, J.~Martinerie, C.~Adam, and F.J. Varela, Physica D {\bf 127}, 250 (1999).

\bibitem{arnhold}
J.~Arnhold, P.~Grassberger, K.~Lehnertz, and C.E. Elger, Physica D
{\bf 134}, 419 (1999).

\bibitem{quian}
R. Quian Quiroga, J. Arnhold and P. Grassberger. Phys. Rev. E, {\bf
  61}, 5142 (2000). 

\bibitem{www}
The three EEG signals to be studied can be downloaded from
  www.physio.mu-luebeck.de/user/rq/data.htm

\bibitem{rosemblum}
M. Rosenblum, A. Pikovsky and J. Kurths. Phys. Rev. Lett, {\bf 76}, 1804 (1996).

\bibitem{lachaux}
J. Lachaux, E. Rodriguez, J. Martinerie and F. Varela. Human Brain Mapping, {\bf 8}, 194 (1999).

\bibitem{tass}
P. Tass, M. Rosenblum, J. Weule, J. Kurths, A. Pikovsky, J. Volkmann,
A. Schitzler and H. Freund. Phys. Rev. Lett, {\bf 81}, 3291 (1998).

\bibitem{zaks}
M. Zaks, E. Park, M. Rosenblum and J. Kurths. Phys. Rev. Lett. {\bf
  82}, 4228, 1999.

\bibitem{florian}
F. Mormann, K. Lehnertz, P. David and C.E. Elger. Physica D, {\bf 144}, 358
(2000).

\bibitem{rodriguez}
E. Rodriguez, N. George, J. Lachaux, J. Martinerie, B. Renault and F. Varela. Nature, {\bf 397}, 430 (1999).








\bibitem{lopes}
F.H. Lopes da Silva, in
{\it Electroencephalography: 
Basic Principles, Clinical Applications and Related Fields}, edited by
E. Niedermeyer and F. H. Lopes da Silva (Baltimore, Williams and
Wilkins 3rd ed., 1993), pp 1097.

\bibitem{takens}
F.~Takens, in D.A. Rand and L.S. Young, eds., {\em Lecture Notes in
  Mathematics 898}, page 366 (Springer, Berlin etc., 1981).

\bibitem{andreas}
A. Schmitz. Phys. Rev. E, {\bf 62}, 7508 (2000). 

\bibitem{grass1}
P. Grassberger. Phys. Lett. A, {\bf 128}, 369, 1988.

\bibitem{thomas}
T. Schreiber. Phys. Rev. Lett. {\bf 85}, 461, 2000.

\bibitem{palus}
M. Palus. Phys. Lett., submitted.

\bibitem{gray}
R. Gray. Entropy and information theory. New York, Springer Verlag,
1990.

\bibitem{kl}
R. Quian Quiroga, J. Arnhold, K. Lehnertz and
P. Grassberger. Phys. Rev. E, {\bf 62}, 8380, 2000.

\bibitem{grass}
P. Grassberger, T. Schreiber and C. Schaffrath. Int.J. of Bifurcation
and Chaos, {\bf 1}, 521, 1991.

\bibitem{grass2}
P. Grassberger. Phys. Lett. A {\bf 97}, 224, 1983.

\bibitem{schuster}
K. Pawelzik and H.G. Schuster. Phys. Rev. A, {\bf 35}, 481, 1987.

\bibitem{theiler}
D. Pritchard and J. Theiler. Physica D {\bf 84}, 476, 1995.


\bibitem{pikovsky} M.G. Rosenblum, A.S. Pikovsky, C. Sch\"afer, P.
  Tass, and J. Kurths.  In: Handbook of Biological Physics; Vol. 4,
  Neuro-informatics (F. Moss and S. Gielen eds.), Elsevier Science, 
  pp. 279-321, 2000.

\bibitem{grossman1}
A. Grossmann, R. Kronland-Martinet and J. Morlet. In: (Combes et
al. eds.) Wavelets: Time-Frequency methods and phase space. Berlin,
Springer (1989).

\bibitem{lachaux1}
J. Lachaux, E. Rodriguez, M. Le van Quyen, A. Lutz, J. Martinerie and
F. Varela. Int. J. Bifurcation and Chaos, {\bf 10}, 2429 (2000).

\bibitem{ott}
D. DeShazer, R. Breban, E. Ott and R. Roy. Phys. Rev. Lett, in press.
 
\bibitem{cohen}
L. Cohen. Time-frequency analysis. Prentice Hall, New Jersey (1995).

\bibitem{giles}
G. van Luijtelaar and A. Coenen (eds.). The WAG/Rij rat model of absence
epilepsy: Ten years of research. Nijmegen University Press, 1997.

\bibitem{drinkenburg}
WHIM Drinkenburg, G. van Luijtelaar, WJ van Schaijk and A
Coenen. Physiol. Behav. {\bf 54}, 779, 1993.

\bibitem{sleepwake} 
G. van Luijtelaar, J. Welting and R. Quian Quiroga. 
In: van Bemmel et al. (eds.) Sleep-wake research in the
  Netherlands, vol 11, pp:86-95. Dutch Society for Sleep-Wake
  Research, 2000.

\bibitem{theiler1}
J. Theiler. Phys. Rev. A, {\bf 34}, 2427 (1986).

\end{thebibliography}

\newpage

\begin{sidewaystable}
\begin{center}
\begin{tabular}[c]{c c c c c c c c c c c c c}
{\bf Example} & {\bf $c_{xy}$} & {\bf $\Gamma_{xy}$} & {\bf $S(R|L)$} & {\bf $S(L|R)$} & {\bf $H(R|L)$} & {\bf $H(L|R)$} & {\bf $N(R|L)$} & {\bf $N(L|R)$}   & {\bf $\gamma_H$} &  {\bf $\gamma_{H-Sh}$} & {\bf $\gamma_W$} &  {\bf
  $\gamma_{W-Sh}$}\\
\hline \\
{\bf A} & 0.70 & 0.88 & 0.34 & 0.34 & 0.67 & 0.60 & 0.46 & 0.42 & 0.59 & 0.12 & 0.71 & 0.19\\
{\bf B} & 0.79 & 0.86 & 0.35 & 0.28 & 1.11 & 1.30 & 0.63 & 0.69 & 0.71 & 0.18 & 0.80 & 0.28\\
{\bf C} & 0.42 & 0.40 & 0.17 & 0.23 & 0.33 & 0.45 & 0.24 & 0.32 & 0.48 & 0.09 & 0.48 & 0.09\\
\end{tabular}
\caption{Synchronization values for the three examples of figure
  \ref{fig:examples}. $c_{xy}$: cross-correlation; $\Gamma_{xy}$:
  coherence (at 9Hz); $S(R|L)$, $H(R|L)$, $N(R|L)$ and
  $S(L|R)$, $H(L|R)$, $N(L|R)$: non-linear
  interdependences of the right electrode on the left and viceversa;
  $\gamma_H$ and $\gamma_{H-Sh}$: phase synchronization indices
  defined from the Hilbert Transform (eq. (\ref{eq:ps}) and
  (\ref{eq:shannonps}), respectively); $\gamma_W$ and $\gamma_{W-Sh}$:
  phase synchronization indices defined from the Wavelet Transform.}
\label{tab:h}
\end{center}
\end{sidewaystable}

\newpage

\newpage

\newpage

\newpage

\newpage

\newpage

\newpage

\newpage
\begin{figure}
\begin{center}
\epsfig{file=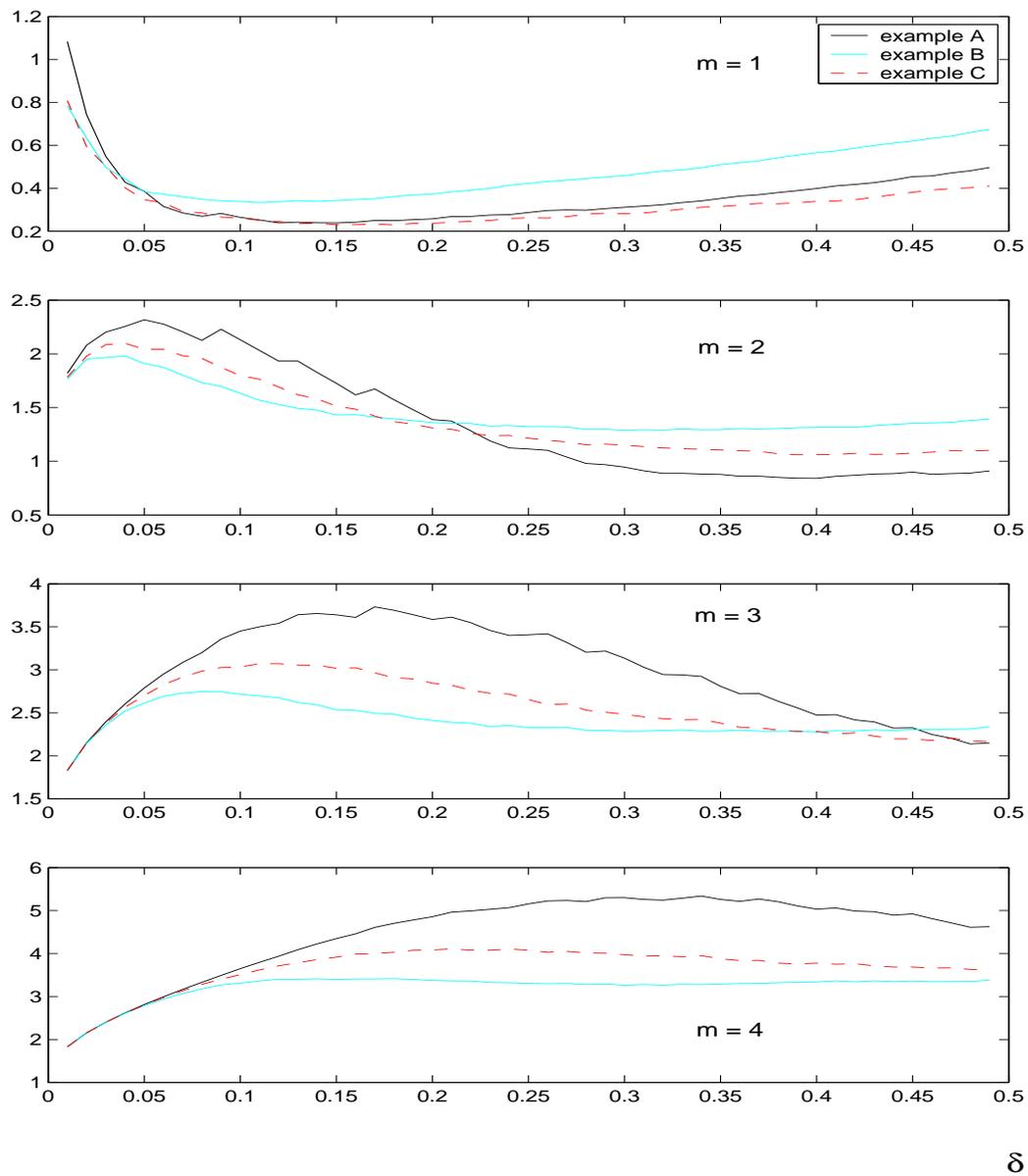, height=16cm,width=14cm,angle=0}
\end{center}
\caption{Mutual information calculated with embedding dimensions $m=1,
  2, 3, 4$ and varying radius  ($\delta$) of the correlation sum.}
\label{fig:mi}
\end{figure}

\end{document}